\documentclass[12pt,preprint]{aastex}

\usepackage{epsfig}

\slugcomment{Submitted to the Astrophysical Journal}

\shorttitle{Cosmic Filaments in Superclusters}
\shortauthors{Bregman, Dupke \& Miller}

\begin{document}

\title{Cosmic Filaments in Superclusters}

\author{Joel N. Bregman, Renato A. Dupke, and Eric D. Miller}
\affil{Department of Astronomy, University of Michigan, Ann Arbor, MI 48109}
\email{jbregman@umich.edu, rdupke@umich.edu, milleric@umich.edu}

\begin{abstract}

Large-scale structure calculations show that modest overdensity filaments
will connect clusters of galaxies and these filaments are reservoirs of
baryons, mainly in gaseous form. To determine whether such filaments exist,
we have examined the UV absorption line properties of three AGNs projected
behind possible filaments in superclusters of galaxies; the AGNs lie within
3 Mpc of the centerlines of loci connecting clusters. \ All three lines of
sight show absorption in Ly$\alpha $, Ly$\beta $, or/and \ion{O}{6} at redshifts
within about 1300 km s$^{-1}$ of the nearby galaxy clusters that
would define the closest filaments. For one AGN, the absorption line
redshifts are close to the emission line redshift of the AGN, so we cannot
rule out self-absorption for this object. \ These absorption line associations 
with superclusters are unlikely to have occurred by chance, a result consistent 
with the presence of cosmic filaments within superclusters.

\end{abstract}


\section{Introduction}

Big-bang nucleosynthesis and quasar absorption line studies (at $z \approx
3$) determine the same mass fraction of baryons in the universe:
$\Omega_{b} = 0.04 h_{75}^{2}$ (e.g., Fukugita, Hogan, and Peebles
1998). It was expected that this gas would become the galaxies of today,
but a census of stars and gas in galaxies in the local universe accounts
for only about 20\% of the baryonic mass. This is the ``missing'' baryon
problem that has drawn much attention, for which the leading solution is
that the baryons have remained as dilute gas (e.g., Fukugita, Hogan, and
Peebles 1998; Cen and Ostriker 1999a; Dav\'{e} et al. 2001). However,
the Ly$\alpha $ forest becomes less dense in the local universe by
about a factor of 3-4 compared to the number of absorption systems at
fixed equivalent width at z = 3.  
This decrease in the baryons is not balanced by the increase in the 
stellar mass, leading to the conclusion that a significant mass of
gas must be hot, above 10$^{5}$ K.

There is a prediction for the properties of the missing baryons from
hierarchical models of structure formation. The most visible large
structures are clusters of galaxies, which are deep potential wells of high
overdensity, but these clusters are relatively rare. The more common
collapsed systems are groups of galaxies and the large filaments that
connect the groups and clusters. Groups of galaxies are weak emitters of
X-rays, indicating gas at 0.3--2$\times 10^{7}$ K but with
extended gaseous masses that are very poorly known because their extent is
broad and their surface brightness fades below the detectability of X-ray
instruments at a fraction of the virial radius.

It is the filaments that contain most of the volume in collapsed structures
and here the temperature is generally lower, typically $10^{5}$--$10^{7}$
K, according to theoretical calculations (e.g., Cen et al. 1995). There
have been no convincing detections of these filaments in emission, but
there is good evidence that gas is present at these temperatures, as
\ion{O}{6} absorption is detected by both HST and FUSE in observations
toward bright AGNs (e.g., Savage et al. 2002). Gas bearing \ion{O}{6}
occurs when $T \sim 10^{5.5}$ K, a temperature expected in filaments, and
because this doublet is strong and oxygen is the most abundant heavy
element, this is an excellent tracer of such gas. These observations show
that \ion{O}{6} absorption is common, $dN/dz \approx 20$ for 
$W_{\lambda} > 50$ m\AA\ (Tripp and Savage 2000). It is predicted that the
\ion{O}{6} absorption lies in a network of filaments with modest
overdensities of 10--40 (Cen et al. 2001) as well as in the galaxy groups
(Hellsten et al.  1998).  When \ion{O}{6} is present, a small fraction of
hydrogen is still neutral, so the Lyman series lines of \ion{H}{1}
accompany \ion{O}{6} as a tracer.

The goal of our effort is to identify the likely locations of these
filaments and then to determine whether \ion{O}{6} and \ion{H}{1} absorption is
associated with them. Theoretical simulations guide us to the most likely
location of filaments (Evrard et al. 2002), which shows that they generally
connect rich clusters, sothe objects with the highest frequency of
filaments (and the largest covering factor) are the superclusters of
galaxies. These are generally several galaxy clusters near each other on
the sky and with only modest velocity differences, so they are physically
close in space. In this study, we examine three AGNs that have already been
observed in the ultraviolet either by the \textit{Far Ultraviolet
Spectroscopic Explorer\/} (\textit{FUSE\/}), or by the \textit{Hubble Space
Telescope\/} (\textit{HST\/}), and each of these show absorption near the
redshift of the supercluster behind which they are projected. We analyze
the properties of these absorption features and discuss the implication for
the detection of the ``cosmic web'' of filaments.

\section{Sample Selection and Results}

Practical aspects of observing, involving the identification of suitably
bright AGNs, drive us toward using the nearby superclusters as the optimum
absorption sites. Nearby superclusters subtend a large solid angle (with
known redshift), which is very helpful for finding sufficiently bright
background AGNs. Also, most of the brightest AGNs are nearby, typically
Seyfert galaxies at z $<$ 0.1, so the superclusters must be at lower
redshift, which are the closest superclusters. Another consideration is that
numerical simuations suggest that filaments connect clusters in a
near-linear fashion, so the background AGN should lie close to loci
connecting clusters. The tangential dimensions of the filaments (at an
overdensity of 10) are predicted to be the same size as the virial radii of
the galaxy clusters ($\approx $ 3 Mpc; Cen and Ostriker 1999b), so
background AGNs should be within this distance of the line connecting
clusters, although, as this is a theoretical prediction, we will include
AGNs that are not so ideally aligned.

Superclusters have been cataloged out to a distance of $z = 0.12$ (Einasto
et al. 1997). They have catalogued 220 superclusters using a
friends-of-friends algorithm with a neighborhood radius of $24 h^{-1}$ Mpc.
We visually identified the most likely locations of filaments connecting
galaxy clusters within superclusters and searched for UV-bright AGNs
projected behind them, using the updated V\'{e}ron-Cetty \& V\'{e}ron
(2001) Catalog and at low Galactic $E(B-V)$ as given by Schlegel,
Finkbeiner, and Davis (1998). About ten objects satisfy our criteria and
are bright enough to be observed by \textit{HST\/} or \textit{FUSE\/} and
we have an observing program to obtain spectra of these targets. However,
three targets have already been observed by \textit{FUSE\/} and
\textit{HST\/} (PHL 1811, PG 1402+261 and TON S180), which we report upon
here.

The locations of these three targets, relative to the foreground
superclusters, are shown in Figure 1, where we indicate the position of
clusters of galaxies and the size of the crosses represent the richness of
the cluster as defined in Dekel \& Ostriker (1999). The value for the
cluster with maximum richness is given and the values for the other
clusters can be estimated by the size of the crosses (in unit steps). Also
shown are the AGNs (empty diamond) and the estimated 2-D location of the
relevant filaments, where the thickness of the filaments is approximately
$3 h^{-1}$ Mpc. We eliminated misclassified clusters from the Einasto et
al. (1997) catalog (those with velocities significantly different from the
rest of the systems in the supercluster; they comprise less than 10\% of
the clusters).  In Table 1 we list the name of the AGN, redshift, visual
magnitude, the supercluster projected in the line-of-sight, its redshift,
the full redshift range of the supercluster, the standard deviation of one
galaxy cluster about the supercluster redshift, and the estimated projected
distance from the AGN to the nearest candidate filament.

\begin{figure}
\begin{minipage}[c]{\linewidth}
\begin{minipage}{.5\linewidth}
\epsfig{file=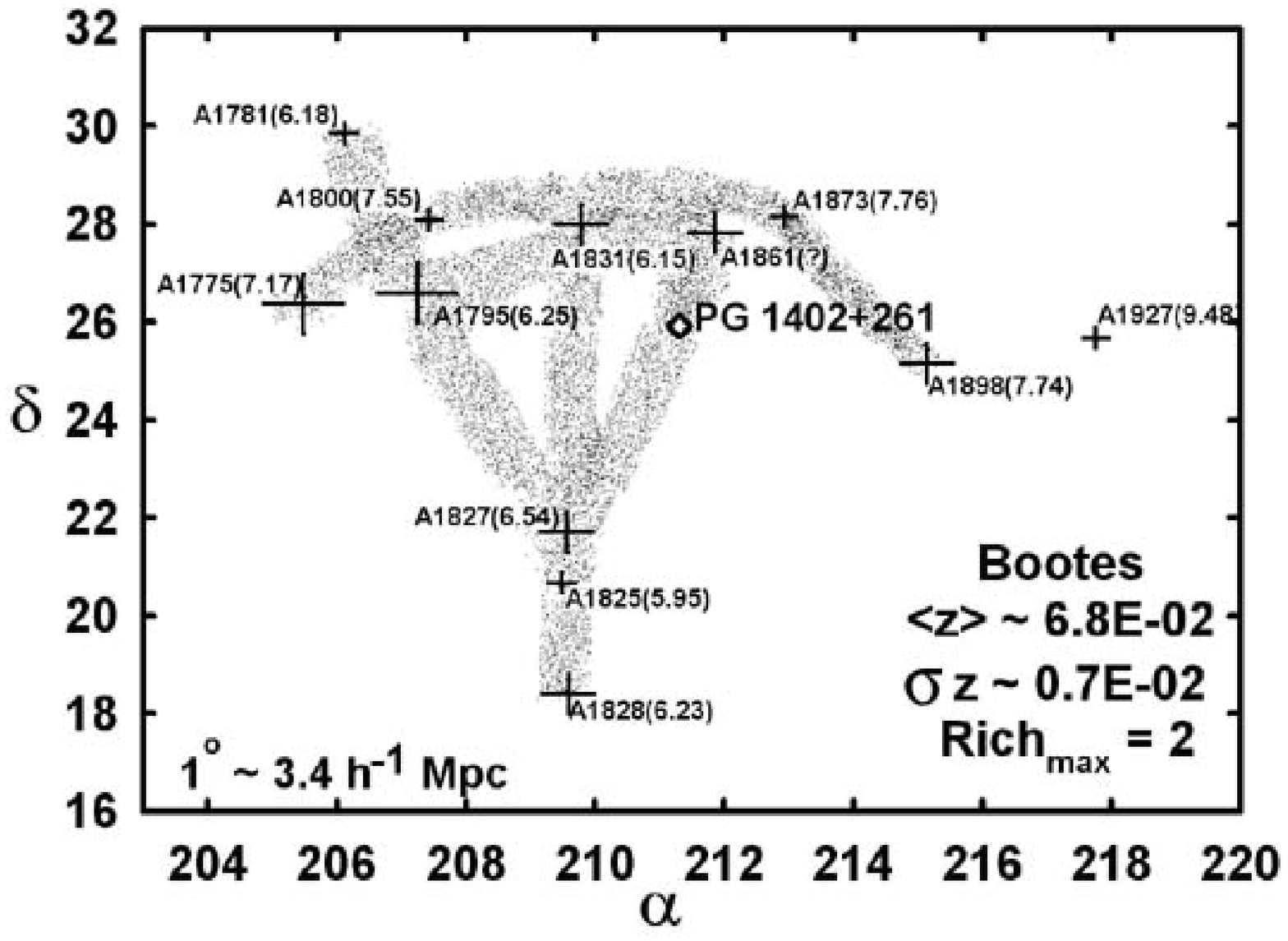,width=\linewidth}
\end{minipage}\hfill
\begin{minipage}{.5\linewidth}
\epsfig{file=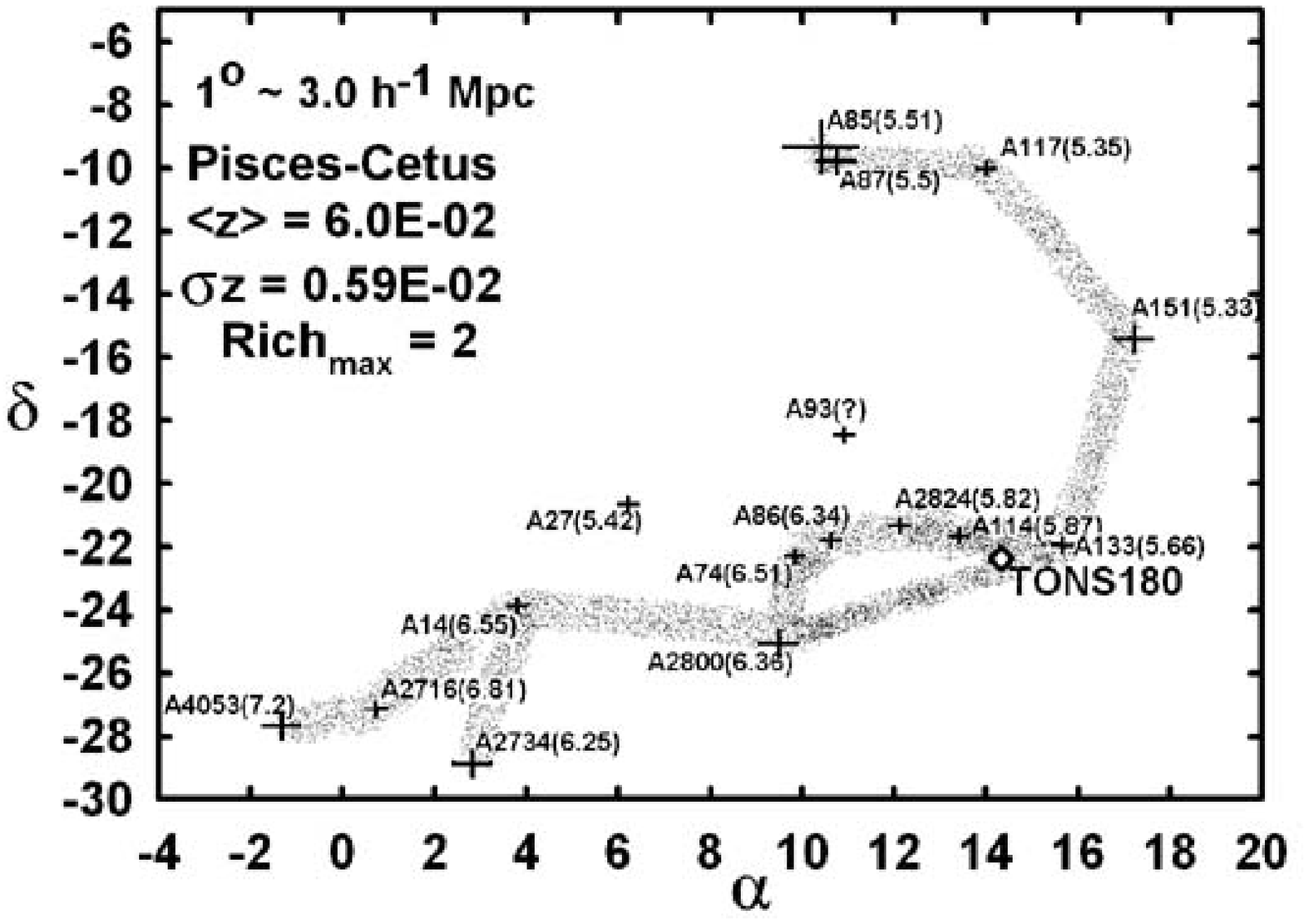,width=\linewidth}
\end{minipage}
\end{minipage}
\begin{minipage}[b]{\linewidth}
\begin{minipage}{.5\linewidth}
\epsfig{file=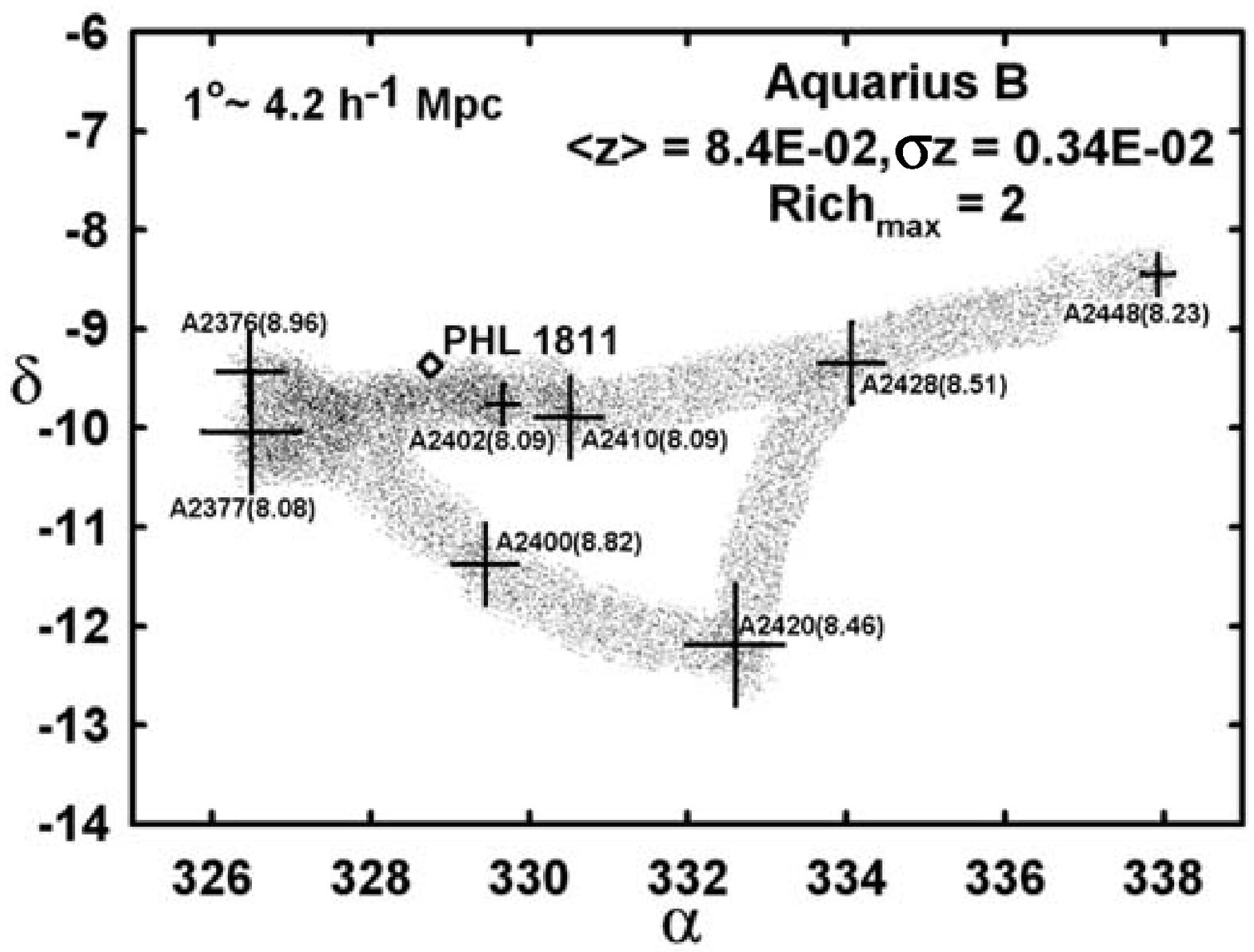,width=\linewidth}
\end{minipage}\hfill
\begin{minipage}{.5\linewidth}
\epsfig{file=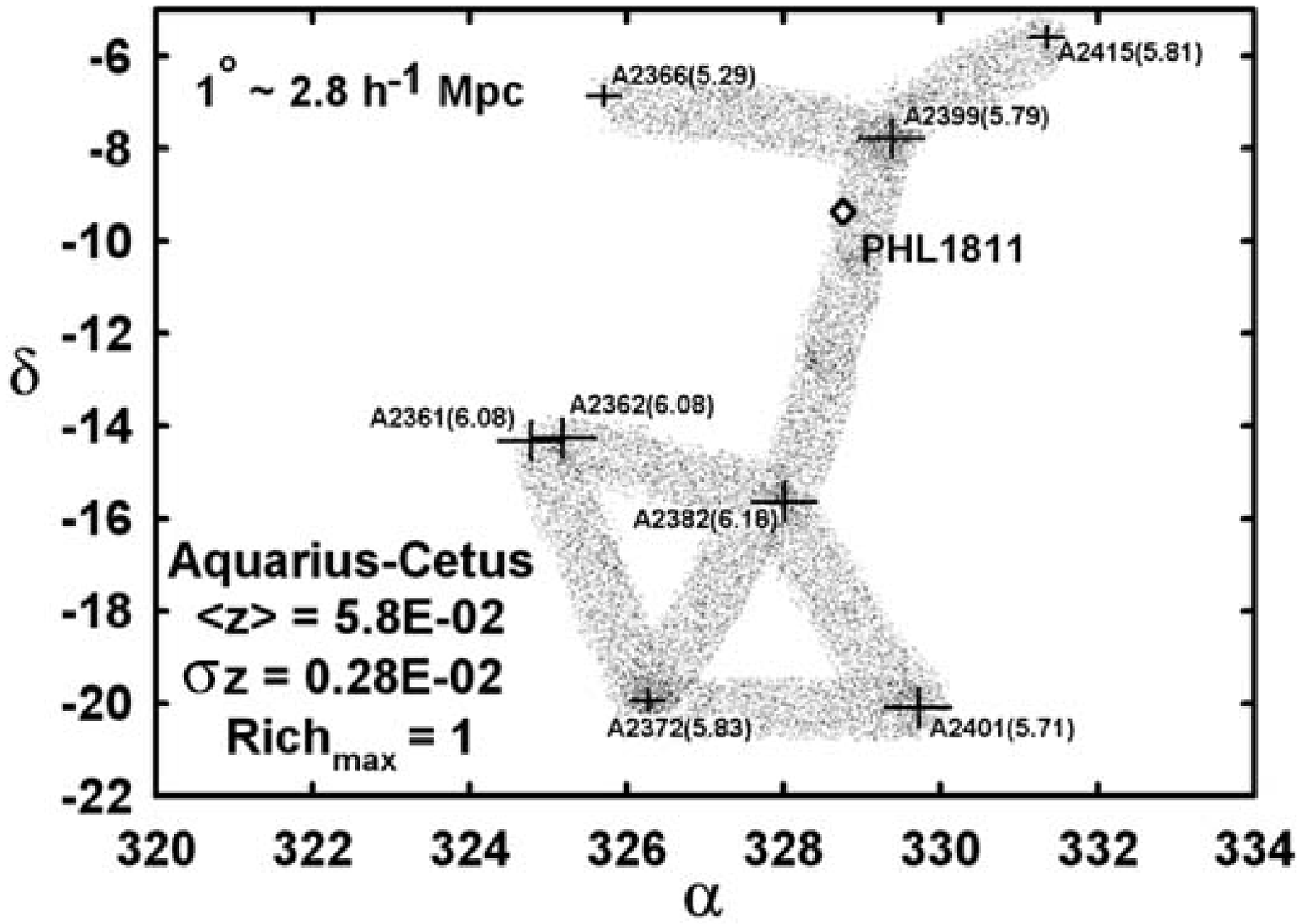,width=\linewidth}
\end{minipage}
\end{minipage}
\caption{Spatial configuration of the background AGNs (open diamonds)
projected onto the superclusters. The crosses are galaxy clusters (many with
Abell names), where the size of the cross indicates the relative population
richness. The linear grey stripes are connections between galaxy clusters
that are possible filaments projected near the background AGN, based upon
three-dimensional proximity and cluster richness; these filaments are 3 Mpc
wide. We also indicate the redshift of the clusters in parenthesis (in units
of 10$^{-2}$), the average redshift of the supercluster, and the standard
deviation of cluster redshifts in the supercluster.
\label{fig1}}
\end{figure}

\begin{deluxetable}{lllllllr}
\tablewidth{0pt}
\tablecaption{Supercluster and AGN Properties\label{tab1}}
\tablehead{
\colhead{Target Name}            &
\colhead{$z_{\rm AGN}$}          &
\colhead{$m_V$}                  &
\colhead{Supercluster Name}      &
\colhead{$z_{\rm SC}$}           &
\colhead{$\Delta z_{\rm SC}$}    &
\colhead{$\sigma(z_{\rm SC})$}   &
\colhead{$\Delta R_{\rm proj}$} 
}
\startdata
PG 1402+261 & 0.164 & 15.5 & Bootes & 0.070 & 0.0151 & 0.007 & $<$ 1 Mpc \\
Ton S180 & 0.062 & 14.4 & Pisces-Cetus & 0.060 & 0.0187 & 0.0059 & $<$ 3 Mpc \\
PHL 1811 & 0.192 & 13.9 & Aquarius-Cetus & 0.058 & 0.0089 & 0.0028 & $<$ 1.5 Mpc \\
PHL 1811 & 0.192 & 13.9 & Aquarius B & 0.084 & 0.0043 & 0.0034 & 2 Mpc \\
\enddata
\end{deluxetable}

The \textit{FUSE} observations were reduced in-house with version 2.2.1 of
the pipeline and reprocessed most recently with version 2.4 of CALFUSE, but
there was no difference in the resulting spectra. \ For analysis purposes,
the \textit{FUSE} spectra were binned by a factor of 5 (0.034 \AA\ width)
and then smoothed with a seven point FFT filter, yielding a resolution of
about 4,000 (we examined the spectra at full resolution as well before
adopting this procedure). For the \textit{HST} data, we used the standard
data products provided by the \textit{MAST}\ archive.

\textbf{PG1402+261}

This Seyfert 1 galaxy, at an emission line redshift of 0.164, lies behind
the Bootes supercluster, which is composed of 11 galaxy clusters in a
redshift range of 0.059-0.079 (there is an additional cluster at $z = 0.0948$
that could be considered an outlier; Figure 1a). This object lies at high
Galactic latitude ($73.4^{\circ}$), with a low extinction ($E(B-V) =
0.016$; Schlegel et al. 1998) and a Galactic neutral hydrogen column of
$1.50 \times 10^{20}$ cm$^{-2}$ (Lockman et al. 2002, Wakker et al.  2003).
There are spectra available for the 900-1180 \AA\ region (\textit{FUSE\/},
a 8 ksec observation taken in June 2000, using the LWRS aperture) as well
as for the 1200-1600 \AA\ region (\textit{HST\/}, with the G130H, G190H,
and G270H on the FOS in August 1996), both of which have been discussed in
the literature (Bechtold et al. 2002; Wakker et al.  2003).

The \textit{FUSE\/} spectrum shows all the strong Galactic atomic
lines, including the high ionization systems (\ion{O}{6}) and the low
ionization systems (\ion{S}{3}, \ion{Si}{2}, \ion{C}{2}, \ion{C}{3},
\ion{O}{1}, \ion{Ar}{1}, \ion{Fe}{2}, \ion{Fe}{3}, and \ion{N}{1}). These
Galactic lines are blueshifted by about 40 km s$^{-1}$, which is probably
due to the target not being located in the center of the slit, but this
shift has no impact on our analysis. There is very little molecular
hydrogen evident, with an upper limit of about 10$^{17.5}$ cm$^{-2}$. In
our analysis of the spectrum, we find no unidentified lines in the
\textit{FUSE} spectral range, aside from the expected number of
2--3$\sigma$ features. In the \textit{HST\/} spectrum, obtained with the
Faint Object Spectrograph, using the G130H spectral element (resolution of
1300, or 230 km s$^{-1}$), there are a few unidentified lines, such as
those noted by Bechtold et al. (2002) at 1383.11 \AA\ and 1405.36 \AA.
These are most likely Ly$\alpha$ absorption features at redshifts of 0.137
and 0.156, respectively.

In addition to the features cataloged by Bechtold et al. (2002),
there is a weak feature at 1285.1 \AA\ that does not correspond to any
Galactic features. It is most likely a Ly$\alpha $ absorption line at a
redshift of 0.0571 and with an equivalent width of 150-200 m\AA\ and an
uncertainty of about 50 m\AA\ (Figure 2). Confirmation of this feature
through the detection of some other Lyman series lines is a commonly used
tool when the wavelength coverage is broad. Unfortunately, nearly all of the
strong lines lie in places in which it is difficult to detect a line
uniquely. That is, the Ly$\beta $ line would lie at 1084.3 \AA , but it
would be confused with the strong Galactic \ion{N}{2} $\lambda $1084.0
line, so a unique idenfitication is not possible. \ Also, the feature does
not fall in the Lif1a channel, but the Sic1a channel, where the S/N is
significantly
lower. \ We estimate that the 3$\sigma $ upper limit to the Ly$\beta $ line
is about 0.25 \AA , which is much larger than the expected value of 0.03 
\AA\ if Ly$\alpha $ were optically thin. \ The Ly$\gamma $ line would be at
1028.1 \AA , in the shoulder of the Galactic Ly$\beta $ absorption line, but
also coincident with an airglow line, that could fill in a weak absorption
feature. \ The line is not detected, with a 3$\sigma $ upper limit of 0.1 
\AA , which is consistent with the predicted equivalent width of 0.01 \AA\ %
if the Lyman lines are thin. \ The Ly$\delta $ line, at 1004.0 \AA , is not
detected but the 3$\sigma $ upper limit, 0.2 \AA , is considerably larger
than the expected line strength. \ We searched for several atomic lines but
detected none (e.g., \ion{C}{3} $\lambda $977, \ion{C}{2} $\lambda $1036). \
For the low ionization lines, this is the expected result, as the Lyman
lines are always stronger, but \ion{O}{6} can be stronger than the Lyman
lines at high temperatures (5$\times 10^{5}$ K). \ The stronger \ion{O}{6}
line (1031.9 \AA ) would fall at 1090.9 \AA , and no line is seen, with a
3$\sigma$ upper limit of 0.1 \AA\ equivalent width.

\begin{figure}
\epsfig{file=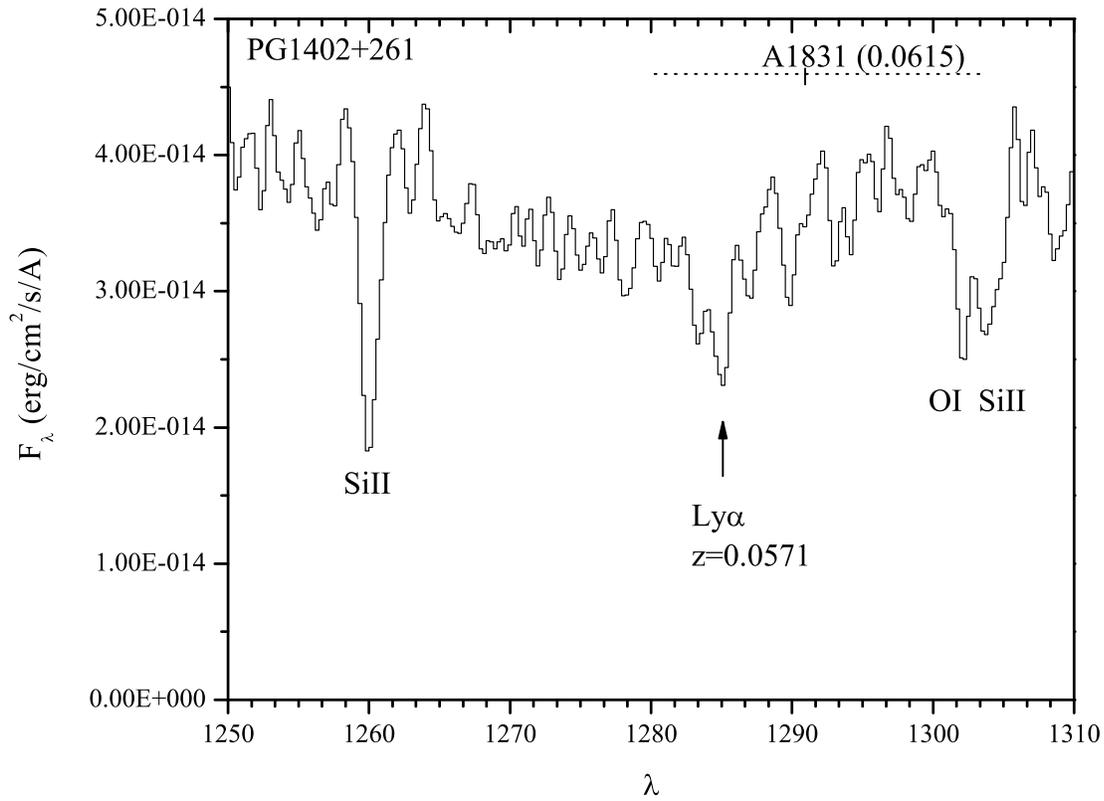,width=\linewidth}
\caption{{\it HST\/}/{\it FOS\/} spectrum of PG1402+261, showing the
strong Galactic lines of \ion{Si}{2} and \ion{O}{1}, as well as an
unidentified absorption feature, which is most likely Ly$\alpha$ near the
redshift of the Bootes supercluster.  The dotted line indicates the region
of expected absorption within $\pm$ 1300 km s$^{-1}$ of the redshift of the
nearest cluster.
\label{fig2}}
\end{figure}

If the line at 1285.1 \AA\ is a Ly$\alpha $ absorption line at $z = 0.0571$,
it is close to the redshift of the supercluster. Of the clusters in the
Bootes supercluster with known redshifts, the one closest to PG1402+261 is
Abell 1831 (2.5$^{\circ }$ away, $z = 0.0615$), which has the smallest
redshift difference with respect to the absorption system, a difference of
1300 km s$^{-1}$, comparable to the velocity dispersion of the cluster.
Consequently, it is possible that this absorption line system is associated
with a supercluster filament that connects to Abell 1831, as it lies beyond
the virial radius of the cluster (a radius of about 0.5$^{\circ } $). There
are other galaxy clusters nearby, but they do not have measured redshifts,
such as Abell 1861, which is closer on the sky to PG 1402+261 than Abell
1831. Two galaxy clusters lie within 1.5$^{\circ }$ of PG1402+261 and are
presumably more distant than the Bootes cluster: Abell 1797 and Abell 1982.
The redshifts of these should be measured in order to determine whether
some of the other unidentified lines could be Ly$\alpha $ absorption
related to the clusters.

\textbf{Ton S180}

This Seyfert 1.2 galaxy lies near the South Galactic Pole (l, b =
139, -85), it has relatively little Galactic absorption (E(B-V) = 0.014),
and it has a redshift of 0.06198. In projection, it lies behind the
Pisces-Cetus supercluster, near two lines of clusters that connect to Abell
133 ($z = 0.0566$; 82$^{\prime }$ away). Along the more northern system are
Abell 114 ($z = 0.0587$; the closest cluster, 66$^{\prime }$ away), Abell 2824
($z = 0.049$; 136$^{\prime }$ away), Abell 86 ($z = 0.061$; 209$^{\prime }$
away), and Abell 74 ($z = 0.0651$; 256$^{\prime }$ away). Toward the south,
Abell 133 may connect to Abell 2800 ($z = 0.0636$; 311$^{\prime }$ away;
Figure 1b). If a filament connects these clusters both spatially and in
velocity, then this Seyfert galaxy is about at the redshift of the filament,
of which it may be part. Being part of a filament increases the likelihood
that our line of sight toward it would pass through overdense regions, but
it also makes it impossible to distinguish between absorption intrinsic to
the AGN or host galaxy and absorption by the filament.

The\ \textit{FUSE} data were obtained in December 1999, using the LWRS
aperture, and with an exposure time of 132 ksec (Shull et al. 2000; Wakker
et al. 2003). \ There were several ultraviolet spectra taken with \textit{%
STIS} on \textit{HST}, using the G140L (January 2000 observation for 1260
seconds) and using the G140M covering the 1194-1300 \AA\ region (taken July
1999) and discussed by Shull et al. (2000), Turner et al. (2002), and Penton
et al. (2004).

The system has the usual set of Galactic atomic absorption lines,
including \ion{O}{6} (Wakker et al. 2003), but it has very little H$_{2}$ 
($< 10^{17}$ cm$^{-2}$). There are several low redshift Ly$\alpha $ and
Ly$\beta $ ``forest'' lines (Shull et al. 2000), but in addition, there are
absorption features near $z = 0.06$ not previously reported. At least three
absorption features are seen in both of the \ion{O}{6} lines in the
redshift range 0.0614 - 0.0625. These redshifts probably need a small
correction upward by about 0.00018 since the Galactic atomic lines are
shifted by that amount (e.g., the nearby \ion{Fe}{2} line, and others), an effect
that can occur if the object is not centered in the slit; the corrected
redshifts are 0.0616, 0.0620, and 0.0625 (Figure 3). The $z = 0.0616$ system
appears to be two lines that are blended together, but the S/N is
insufficient to be certain. The other lines appear single and have the same
line width, a FWHM of 150$\pm $40 m\AA .

\begin{figure}
\epsfig{file=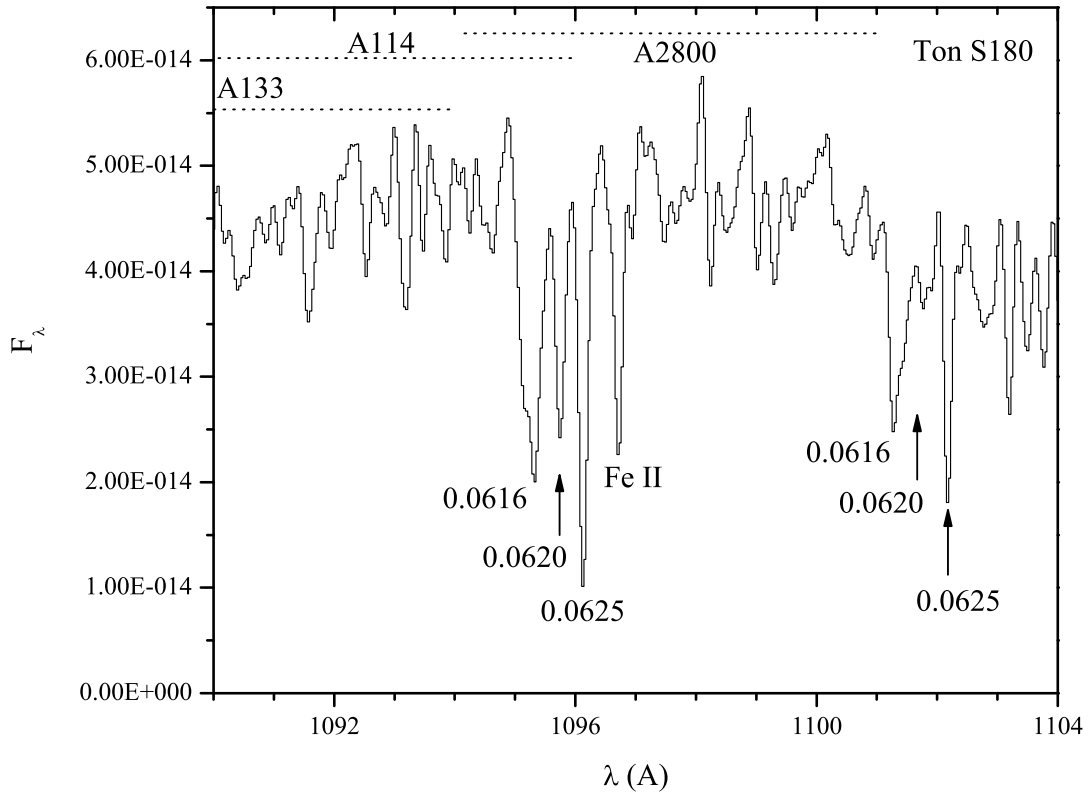,width=\linewidth}
\caption{{\it FUSE\/} spectrum of Ton S180 shows three \ion{O}{6}
absorption line systems near the redshift of the Pisces-Cetus supercluster
as well as the emission line redshift of Ton S180. The lines have a FWHM of
0.15 \AA, which is close to the thermal Doppler width for gas at 
$3 \times 10^5$ K (0.11 \AA).  The dotted line indicates the region of
expected absorption within $\pm$ 1300 km s$^{-1}$ of the redshift of the
nearest cluster.
\label{fig3}}
\end{figure}

These absorption systems also appear in the STIS medium resolution
spectrum of this object with a resolution of about 1000 (Turner et al.
2002), where three Ly$\alpha $ absorption systems are seen at $z = 0.0615$,
0.0620, and 0.0623, with equivalent widths of 80 m\AA , 30 m\AA , and 100
m\AA. \ If these lines are optically thin, the Ly$\beta $ lines would be 13
m\AA , 5 m\AA , and 16 m\AA , but the 1$\sigma $ uncertainty for a line in
this part of the Lif2a spectrum is about 30 m\AA , so the failure to detect
the Ly$\beta $ line is suggestive that the lines are of low optical depth.
\ These Ly$a$ lines are similar to but not exactly at the same redshift,
with differences up to 60 km s$^{-1}$ for the highest and lowest redshift
systems.

The origin of these four redshift systems is ambiguous because we cannot
distinguish between absorption intrinsic to the Seyfert and hot material
within a cosmological filament. If the absorption is associated with a
filament from a cluster, the lines are 800-1050 km s$^{-1}$ away from the
redshift of the nearest cluster (Abell 114) and may be associated with this
system (the line of sight passes about two virial radii from the cluster
center). The properties of the absorption systems are consistent with a hot
but quiescent medium. The equivalent widths of the \ion{O}{6} lines exceed
that of the Ly$\alpha $ lines, typically by factors of 2-3, which can occur
in collisionally ionized gas in the 2--10$\times 10^{5}$ K range (Tripp and
Savage 2000), the type of material predicted to lie in filaments. The line
widths of the \ion{O}{6} lines are close to the thermal width of oxygen at
3$\times 10^{5}$ K (110 m\AA ), so the gas does not lie in a turbulent
medium, as the sound speed of a gas at this temperature is about 
80 km s$^{-1}$ and if the turbulent speed were equal to the sound speed, it
would lead to a FWHM of 500 m\AA\ in all lines.

\textbf{PHL 1811 }(J2155-0922)

This narrow-line Seyfert 1 galaxy, at a redshift of 0.192, is a a
very bright object, although at a galactic latitude of -44$^{\circ }$, there
is more Galactic extinction along this line of sight, with E(B-V) = 0.046.
It lies behind the Aquarius B supercluster as well as the Aquarius-Cetus
supercluster, which are distinguished by their redshifts. In the Aquarius B
supercluster, the cluster closest to PHL 1811 is Abell 2402 (50$^{\prime }$;
$z = 0.0809$), followed by Abell 2410 (109$^{\prime }$; $z = 0.0809$), Abell
2376 (136$^{\prime }$; $z = 0.0896$), Abell 2377 (139$^{\prime }$; $z =
0.0808$), and Abell 2400 (126$^{\prime }$; $z = 0.0882$). If filaments connect
the clusters and pass in front of PHL 1811, the most natural connections
would be between Abell 2376/2377 and Abell 2402/2410 (Figure 1c). \ It has
been observed with HST on several occassions with \textit{STIS} (with the
G140L in December 2001, and in a current program with the E140M), and with 
\textit{FUSE} (in June 2001, and for longer exposures in a current program).%

This system has been studied exhaustively by Jenkins et al. (2003),
where they combine \textit{FUSE\/} and \textit{HST\/} data, permitting them
to determine the strength of many Galactic lines as well as intervening
systems. Four intervening absorption line systems are identified at a
redshift below 0.1, and they are at $z =$ 0.07344, 0.07779, 0.07900, and
0.08093 (Figure 4; the highest redshift system is a Lyman limit system with
17.5 $<$ logN(HI) $<$ 19.5). Of these systems, they identify many Lyman
lines and a few metal lines, including \ion{O}{6} in the $z = 0.07779$ and 
$z = 0.08093$ systems at or above the 3$\sigma $ level, several times
weaker than Ly$\beta $. Jenkins et al. (2003) argue that the Lyman limit
system is associated with an L$_{\ast}$ galaxy that is 34 kpc
from the line of sight and at the same redshift. The origin of the other
absorption systems is less clear, as the velocity differences are great
enough (580, 940, and 2250 km s$^{-1}$) that the gas is unlikely
to be gravitationally bound to the galaxy. They suggest that some of this
absorbing material is due to tidal interactions between galaxies.

\begin{figure}
\epsfig{file=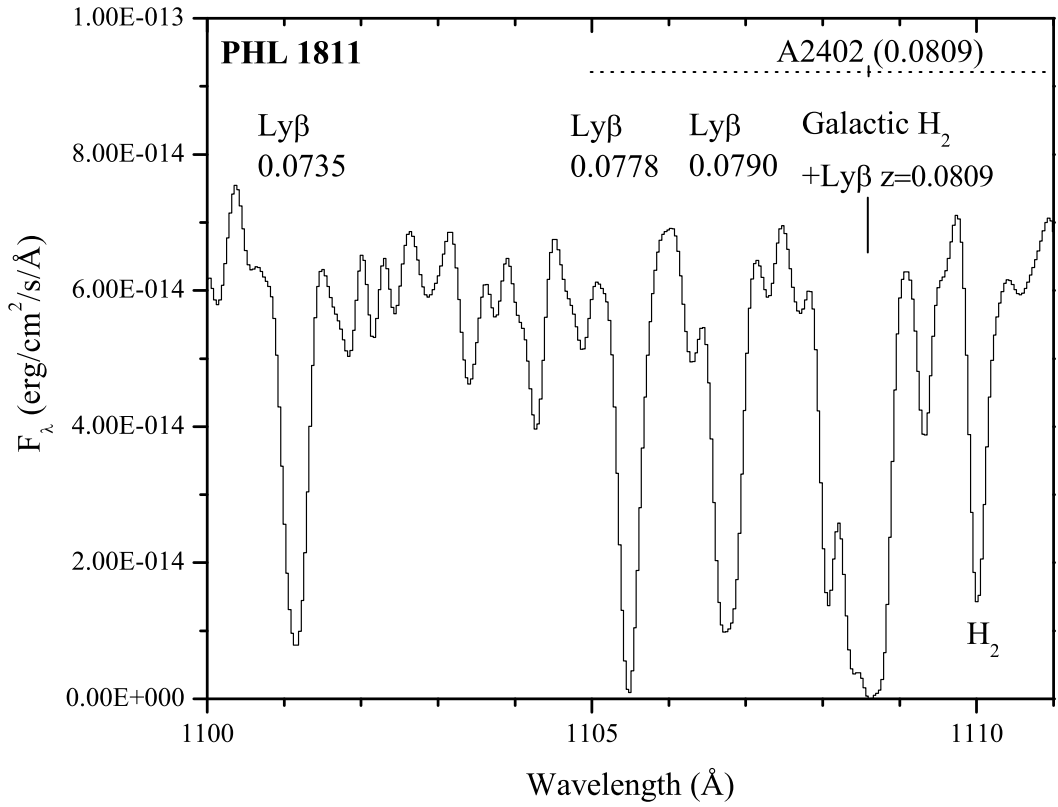,width=\linewidth}
\caption{{\it FUSE\/} spectrum of PHL 1811, showing four Ly$\beta$
systems near the redshift of the Aquarius B supercluster (plus two Galactic
H$_2$ lines).  The dotted line indicates the region of expected absorption
within $\pm$ 1300 km s$^{-1}$ of the redshift of the nearest cluster.
\label{fig4}}
\end{figure}

We have little to add to this analysis except to point out that it is
possible that three systems are due to absorption within the supercluster.
Aside from the strongest stystem at $z = 0.08093$, the next two systems
closest in redshift space are within 940 km s$^{-1}$, while the
most distant system is 2250 km s$^{-1}$ away, which is still
within the velocity range of a supercluster (or possibly even a cluster).
The line of sight toward PHL 1811 is at least twice the virial radius 
from the nearest cluster (projected separation of 50$^{\prime }$; the virial
radius would be about 25$^{\prime }$), so the absorbing material is not
gravitationally bound to any of the clusters. \ Despite an extensive study
by Jenkins et al., they do not identify any galaxies with redshifts that
might be associated with these other redshift systems, nor are there tidal
features evident with the Lyman limit galaxy.

Also, this line of sight passes through the Aquarius-Cetus
supercluster and it lies closest to Abell 2399 ($z = 0.0579$; 102$^{\prime }$
away) and Abell 2366 ($z = 0.0529$; 236$^{\prime }$), with a few clusters to
the south, such as Abell 2361 and Abell 2362 ($z = 0.0608$), although these
are more than 6$^{\circ }$ away (Figure 1d). Jenkins et al. (2003) report a
Ly$\alpha $ absorption line system at a redshift of 0.05119 (equivalent
width of 0.60$ \pm $0.05 \AA ), a redshift difference of 0.0067 (2000 km
s$^{-1}$) for Abell 2399 and 0.0017 (500 km s$^{-1}$)
for Abell 2366 (Figure 4). This is consistent with absorption by material
in this second supercluster along the line of sight.

\section{Summary and Concluding Remarks}

We identified three bright AGNs that have been observed by \textit{FUSE }and%
\textit{\ HST} and lie near possible supercluster filaments. \ Absorption
lines, most of which were reported previously, are found to exist at
redshifts within about 1300 km sec$^{-1}$ of the supercluster, often being
closest in redshift to the nearest galaxy cluster. \ These coincidences
between the supercluster redshifts (or the redshifts of the nearest
clusters) and the absorption line systems are important only if they are
statistically significant, so we need to calculate the probability that this
has occurred by chance. The redshift frequency of Ly$\alpha $ absorption
lines was the topic of an HST Key project as well as more recent work
(Dobrzycki et al. 2002 and references therein), where at low redshift, the
frequency of Ly$\alpha $ absorption lines with equivalent widths greater
than 0.24 \AA\ is dN/dz $\approx $28 (for EW $>$ 0.36 \AA , dN/dz $\approx $%
15). We do not have Ly$\alpha $ equivalent widths for all of the systems
discussed above because in some, only Ly$\beta $ was detected without
blending, as for the $z = $0.077--0.081 absorption line systems in PHL1811.
However, if the ratio of Ly$\beta $/Ly$\alpha $ is about 0.5, as found for
other systems of similar equivalent widths (Shull et al. 2000), then the Ly$%
\alpha $ equivalent widths are above the 0.24 \AA threshold (probably about
0.4 \AA ) and the sum of the three lines have an equivalent width of 1.73 
\AA , consistent with this estimation (Jenkins et al. 2003). For Ton S180,
the absorption is due to \ion{O}{6} but it is challenging to detect Ly$\alpha $
absorption as it would lie in the Ly$\alpha $ emission line, but regardless
of where one might choose a sensible continuum, there is no Ly$\alpha $
absorption line with an equivalent width above 0.05 \AA . However, as the
\ion{O}{6} clearly shows evidence for absorption, we will consider the effect of
including this system in the analysis. Finally, the absorption line from PG
1402+261 is slightly below the threshold, so we will consider the results of
the analysis with and without this system.

The next issue is how one counts a system and over what velocity
width. The most conservative approach is to count multiple absorption line
systems as a single absorption system (i.e., the 0.077-0.081 systems in
PHL1811 and the 0.0616 - 0.0630 systems in Ton S180) even though the
determination for dN/dz includes all redshift systems. The absorption line
systems discussed above fall within $\pm $1300 km s$^{-1}$ ($\pm$ 0.0043 in
redshift space) of the expected redshift of the filament or nearby
cluster, so we should expect 0.24 absorption systems per supercluster. If we
use only PHL 1811, there are absorption systems associated with two
superclusters (excluding the $z = 0.08093$ system and counting the systems at
$z = 0.07779$ and 0.07900 as a single system), which should occur by chance
4.6\% of the time (for dN/dz $\approx $28). Using the rate for absorption
system equivalent widths above 0.36 \AA\ (appropriate for these absorption
systems; dN/dz $\approx $15), the chance probability is 1.5\%. If we include
the absorption system from PG1402+261, the probability of this occurring by
chance falls to 1.4\% (for dN/dz $\approx $28; if we were to include the
\ion{O}{6}
absorption from Ton S180 as a single system, the chance probability would
become 0.3\%).

When we include all of the absorption systems separately, the association
between superclusters and absorption features becomes more likely, as there
would be three absorption systems in the redshift range 0.077-0.081 in
PHL1811 and three in the redshift range 0.0616 - 0.0630 in Ton S180. \
Excluding Ton S180, the probability that the association with the
superclusters occurred by chance is 9$\times 10^{-5}$ (1.1$\times 10^{-3}$
excluding the $z = 0.08093$ system), and when Ton S180 is included, the
probability becomes 1.7$\times 10^{-7}$ (2.2$\times 10^{-6}$ excluding the z
= 0.08093 system).

It is difficult at this point to demonstrate that the absorption is
associated with regions of mild overdensities (filaments) rather than
material that has been ejected or torn away from a galaxy. One would need to
be able to determine if absorbing material is gravitationally bound to the
galaxy, which would involve assuming a model for the galaxy as well as for
the transverse velocity of the absorbing gas relative to the galaxy. Also,
the distinction between these two associations may be unimportant, as
material expelled in a galactic wind is unbound to that galaxy and becomes
part of the filament once it mixes with this surrounding medium.

An alternative approach to looking for evidence of absorption from galaxy
clusters and superclusters is to choose bright but randomly located AGNs and
identify the absorption sites by obtaining imaging and spectroscopy of
nearby objects. Along these lines, Penton, Stocke, and Shull (2002) found
that about 22\% of absorbers are not associated with galaxies (also, McLin
et al. 2002). They find support that absorption line systems align with
large-scale filaments of galaxies, which may be similar in nature to the
filaments that we have tried to probe.

The association of absorption systems with supercluster filaments or
galaxy clusters may seem surprising since a dedicated search to find
absorbing gas in galaxy clusters failed to produce detections (Miller,
Bregman, and Knezek 2002). However, that study was sensitive only to low
ionization metal lines, not to the more common Ly$\alpha $ absorption line
systems (or to \ion{O}{6}). Given the results here, a search to identify Ly$\alpha $
systems around galaxy clusters would be worthwhile.

The relationship between superclusters and Ly$\alpha $ absorption systems is
enticing, but it is based on the study of just three systems, and it should
be possible to improve the statistical situation quite readily. In the near
future, we will be investigating several other low redshift superclusters (z 
$<$ 0.12) to search for Ly$\alpha $ absorption systems. Another observation
that is badly needed is to obtain redshifts for the clusters that are near
the lines of sight. We hope that studies such as those will lead to a deeper
understanding of whether modest overdensity cosmological filaments can be
identified through Ly$\alpha $ absorption line studies.

After this paper was submitted, it was pointed out to us that there is
evidence of a Ly$\alpha $ absorber associated with the supercluster recently
discovered from the \textit{ROSAT} North Ecliptic Pole Survey (NEP). The NEP
supercluster (Mullis et al. 2001) is at an average redshift of 0.087 and is
X-ray selected. It subtends a large angular region, where about 200
AGNs are found. One of them is bright enough for absorption line studies and
was previously observed by HST, H1821+643 (NEP5340). Spectral analyses of
H1821+643 (z=0.297) revealed the presence of a Ly$\alpha $ system at $z =
0.0891$ (Tripp, Lu \& Savage 1998). The nearest galaxy cluster is at a
projected distance of 18 h$^{-1}$ Mpc, well beyond the virial radius; no
galaxy is identified near the absorption site. \ These observations are
consistent with the absorber being associated with diffuse gas connecting
the NEP clusters. \ As our selection criteria did not include this
supercluster, we do not include it in our statistical analysis, but it
supports our findings for the other superclusters.

We would like to acknowledge support for this work from NASA grants
NAG5-10765 and NAG5-10806, which support research on archival data analysis.
Also, we would like to thank B-G Andersen, Ken Sembach, Bart Wakker, Jimmy
Irwin, Chris Mullis, and Edward Lloyd-Davies for their advice and
encouragement.

\end{document}